\documentclass{article}
\usepackage{a4}
\usepackage{psfig}
\usepackage{amsmath}
\usepackage{bm}
\begin{document}
\title{Strong-coupling effects in the relaxation dynamics of ultracold
neutral plasmas}

\author{T.\ Pohl and T.\ Pattard}
\maketitle
{Max Planck Institute for the Physics of Complex Systems,\\
N\"othnitzer Str.\ 38, D-01187 Dresden, Germany}

\begin{abstract}
We describe a hybrid molecular dynamics approach for the description
of ultracold neutral plasmas, based on an adiabatic treatment of the electron
gas and a full molecular dynamics simulation of the ions, which allows us to
follow the long-time evolution of the plasma including the effect of the
strongly coupled ion motion. The plasma shows a rather complex relaxation
behavior, connected with temporal as well as spatial oscillations of the
ion temperature. Furthermore, additional laser cooling of the
ions during the plasma evolution drastically modifies the expansion dynamics,
so that crystallization of the ion component can occur in this nonequilibrium
system, leading to lattice-like structures or even long-range order resulting
in concentric shells.
\end{abstract}

\section{Introduction}
Experiments in cooling and trapping of neutral gases have paved the way
toward a new parameter regime of ionized gases, namely the regime of
ultracold neutral plasmas (UNPs). Experimentally, UNPs are produced by
photoionizing a cloud of laser-cooled atoms collected in a magneto-optical
trap \cite{Kil99}, with temperatures down to 10 $\mu$K. By tuning the frequency
of the ionizing laser, initial electron kinetic energies of $E_{\rm{e}}/k_{\rm B} = 1
\mbox{K} - 1000 \mbox{K}$ have been achieved. The time evolution of several
quantities characterizing the state of the plasma, such as the plasma
density \cite{Kil99,Kul00,Sim04}, the rate of expansion of the plasma cloud
into the surrounding vacuum \cite{Kul00}, the energy-resolved population of
bound Rydberg states formed through recombination \cite{Kil01}, or electronic
\cite{Rob04,Van04} as well as ionic \cite{Sim04} temperature have been measured
using various plasma diagnostic methods.

Despite the low typical densities of $\approx 10^9$ cm$^{-3}$, the very low
initial temperatures suggest that these plasmas have been produced well within
the strongly coupled regime, with Coulomb coupling parameters up to
$\Gamma_{\rm{e}} = 10$ for the electrons and even $\Gamma_{\rm{i}} = 30000$ for the ions.
Thus, UNPs seem to offer a unique opportunity for a laboratory study of neutral
plasmas where, depending on the initial electronic kinetic energy, either one
component (namely the ions) or both components (ions and electrons) may be
strongly coupled. Moreover, the plasma is created in a completely
uncorrelated state, i.e.\ far away from thermodynamical equilibrium. The
relaxation of a strongly correlated system towards equilibrium is an
interesting topic in non-equilibrium thermodynamics and has been studied for
decades. The history of this problem must be traced back to the important 
contributions of Klimontovich
\cite{Kli72,Kli73,Kli82}, who pointed out that kinetic energy conserving
collision integrals such as the Boltzmann, Landau and Lenard-Balescu
collision integrals are not appropriate for such a situation, and derived
non-Markovian kinetic equations taking correctly into account total energy
conservation of the system. In the following years this problem has attracted
much attention and the relaxation of nonequilibrium strongly coupled plasmas 
has been studied by a variety of different methods
\cite{Wal78,Bel96,Bon98,Zwi99}. 
The very low densities of UNPs make it now possible to directly
observe the dynamical development of spatial correlations, which may
serve as the first experimental check of the present understanding of the
strongly coupled plasma dynamics. Moreover, it turns out that the timescale of
the plasma expansion, the correlation time as well as the relaxation time of
the ions are almost equal. Therefore Bogoliubov's functional hypothesis,
usually used in kinetic theory, breaks down under the present conditions,
which may lead to a very interesting relaxation behavior but also causes some
difficulties in the theoretical description of these systems, since the
plasma dynamics can not be divided into different relaxation stages.

\section{Theoretical approach}
A full molecular dynamics simulation of ultracold neutral plasmas over
experimentally relevant timescales is infeasible with present-day computer
resources due to the large number of particles ($N \approx 10^5$) and the
long observation times ($t \approx 10^{-4}$ s) involved.
In order to model the evolution of UNPs, we have developed a hybrid molecular
dynamics (HMD) approach which treats electrons and ions on different levels
of sophistication, namely in a hydrodynamical approximation on the one hand
(for the electrons) and on a full molecular dynamics level on the other hand
(for the ions) \cite{PPR04}. For the electrons, it has been shown that several
heating effects, such as continuum threshold lowering \cite{Hah02}, build-up
of correlations \cite{Kuz02}, and, predominantly, three-body recombination
\cite{Rob02} rapidly increase the electronic temperature. As a consequence,
the electrons are always weakly coupled, $\Gamma_{\rm{e}} < 0.2$, over the whole
course of the system evolution. Moreover, due to the small electron-to-ion
mass ratio, the relaxation timescale of the electrons is much smaller than
that of the ions as well as the timescale of the plasma expansion. Hence, an
adiabatic approximation may safely be applied, assuming instant equilibration
of the electrons. This allows for the use of much larger timesteps than in
a full MD simulation since the electronic motion does not need to be
resolved. It is this adiabatic approximation for the electrons which makes
a molecular dynamics treatment of the ionic motion in UNPs computationally
feasible. The electronic density is determined self-consistently from the
Poisson equation. The fact that the potential well created by the ions which is
trapping the electrons has a finite depth is taken into account by using a
King-type distribution \cite{Kin66} known from simulations of globular clusters
rather than a Maxwell-Boltzmann distribution for the electron velocities, with the electronic temperature $T_{\rm{e}}$ obtained from energy
conservation. The finite well depth also leads to evaporation of a
fraction of the free electrons in the very early stage of the plasma evolution,
which is accounted for by determining the fraction of trapped electrons
from the results of \cite{Kil99}. The dynamics of the heavy particles is
described in the framework of a chemical picture, where inelastic processes,
namely three-body recombination and electron impact
ionization, excitation and deexcitation, are taken into account on the basis
of Boltzmann-type collision integrals \cite{Kli81,Kli82}, with the transition
rates taken from \cite{Man69}. Numerically, the resulting collision integrals
are evaluated using a Monte Carlo sampling as described in
\cite{Rob03,PPR04,PPR04c}. The ions and recombined atoms are then propagated
individually in a molecular dynamics simulation, taking
into account the electronic mean-field potential and the full interaction
potential of the remaining ions\footnote{In order to bring out clearly the role
of ionic correlations, it is also possible to neglect them in the HMD approach
by propagating the ions in the mean-field
potential created by all charges rather than the full ionic interaction.}.
In order to allow for larger particle numbers, the most time-consuming part
of the HMD simulation, namely the
calculation of the interionic forces, is done using a treecode procedure
originally designed for astrophysical problems \cite{Bar90}, which scales like
$N_{\rm{i}} \ln N_{\rm{i}}$ rather than $N_{\rm{i}}^2$ with the number $N_{\rm{i}}$ of ions. 

As shown in several publications \cite{PPR04,PPR04a,PPR04b,PPR04c}, the HMD
approach outlined above provides a powerful method for the description of
UNPs, taking full account of ionic correlation effects. However, due to the
large numerical effort involved, it is limited to particle numbers of $N_{\rm{i}}
\approx 10^5$. While this permits a direct simulation of many, particularly
of the early, experiments, an increasing number of experiments is performed
with larger particle numbers up to $10^7$. Thus, an alternative method which
is able to treat such larger systems is desirable. Such a method is indeed
available \cite{PPR04}, based on a hydrodynamical description of both electrons
and ions similar to that introduced in \cite{Rob02,Rob03}. Starting from the
first equation of the BBGKY hierarchy, one obtains the evolution equations for
the one-particle distribution functions $f$ of the electrons and ions.
Neglecting again electron-electron as well as electron-ion correlations, and
employing the same adiabatic approximation for the electrons already used in
the HMD approach, a quasineutral approximation \cite{Dor98} permits
expressing the mean-field
electrostatic potential in terms of the ionic density, leading to a closed
equation for the ion distribution function which contains the electron
temperature as a parameter. A Gaussian ansatz for the ion distribution
function,
\begin{equation} \label{e1}
f_{\rm{i}} \propto \exp{\left(-\frac{r^2}{2\sigma^2}\right)}\exp{\left(-
\frac{m_{\rm{i}}\left({\bf{v}}-\gamma{\mathbf{r}}\right)^2}{2k_{\rm{B}}
T_{\rm{i}}}\right)} \; ,
\end{equation}
which corresponds to the initial state of the plasma cloud, is then inserted
into the evolution equations for the second moments $\langle r^2 \rangle$,
$\langle \bf{r} \bf{v} \rangle$ and $\langle v^2 \rangle$ of the ion
distribution function. In this way,
evolution equations for the width $\sigma$ of the cloud, the parameter $\gamma$
of the hydrodynamical expansion velocity $\bf{u} = \gamma{\mathbf{r}}$ and
the ionic temperature $T_{\rm{i}}$ are obtained. Ionic correlations are taken into
account in an approximate way using a local density approximation together
with a gradient expansion, reducing the description of their influence on the
plasma dynamics to the evolution of a single macroscopic quantity, namely the
correlation energy $U_{\rm{ii}}$ of a homogeneous plasma. The relaxation behavior
of $U_{\rm{ii}}$ is modeled using a correlation-time approximation \cite{Bon96}
with a correlation time equal to the inverse of the ionic plasma frequency,
$\tau_{\rm corr} = \omega_{\rm{p,i}}^{-1}$, together with an analytical expression
for the equilibrium value of $U_{\rm{ii}}$ \cite{Cha98}. Finally, inelastic
processes such as three-body recombination and electron impact ionization,
excitation and deexcitation are incorporated on the basis of rate
equations, and the influence of the recombined Rydberg atoms on the expansion
dynamics is taken into account assuming equal hydrodynamical velocities for
atoms and ions. The final set of evolution equations then reads
\begin{subequations} \label{e2}
\begin{eqnarray}
\label{e2a}
\dot{\sigma}&=&\gamma\sigma\;,\\
\label{e2b}
\dot{\gamma}&=&\frac{N_{\rm{i}}\left(k_{\rm{B}}T_{\rm{e}}+k_{\rm{B}}T_{\rm{i}}+
\frac{1}{3}U_{\rm{ii}}\right)}{\left(N_{\rm{i}}+N_{\rm{a}}\right)m_{\rm{i}}
\sigma^2}-\gamma^2\;,\\
\label{e2c}
k_{\rm{B}}\dot{T}_{\rm{i}}&=&-2\gamma k_{\rm{B}}T_{\rm{i}}-\frac{2}{3}\gamma
U_{\rm{ii}}-\frac{2}{3}\dot{U}_{\rm{ii}}\;,\\
\label{e2d}
\dot{U}_{\rm{ii}}&=&-\omega_{\rm{p,i}}\left(U_{\rm{ii}}-U_{\rm{ii}}^{\rm{(eq)}}
\right)\\
\label{e2e}
\dot{{\cal{N}}}_{\rm{a}}(n)&=&\sum_{p\neq n}\left[R_{\rm{bb}}{(p,n)}
{\cal{N}}_{\rm{a}}(p)-R_{\rm{bb}}{(n,p)}{\cal{N}}_{\rm{a}}(n)\right]
\nonumber\\&&+R_{\rm{tbr}}(n)N_{\rm{i}}-R_{\rm{ion}}{(n)}{\cal{N}}_{\rm{a}}(n)
\end{eqnarray}
and the electronic temperature is determined by energy conservation,
\begin{equation}
\label{e2f}
N_{\rm{i}}k_{\rm{B}}T_{\rm{e}}+\left[N_{\rm{i}}+N_{\rm{a}}\right]
\left[k_{\rm{B}}T_{\rm{i}}+m_{\rm{i}}\gamma^2\sigma^2\right]+\frac{2}{3}
N_{\rm{i}}U_{\rm{ii}}-\frac{2}{3}\sum_n{\cal{N}}_{\rm{a}}(n)
\frac{{\cal{R}}}{n^2}={\rm{const.}}\;,
\end{equation}
\end{subequations}
where ${\cal{N}}_{\rm{a}}(n)$ defines the population of Rydberg states,
$N_{\rm{a}}=\sum_n{\cal{N}}_{\rm{a}}(n)$ is the total number of atoms and
${\cal{R}}=13.6$eV is the Rydberg constant.

The preceeding hydrodynamical method is much more approximate than the HMD
approach, but, on the other hand, it is much simpler and quicker.
For particle numbers of $N_i \approx 10^5$, it requires about two orders
of magnitude less CPU time. Since its computational effort is independent of
the number of particles, it allows for a simulation of larger plasma clouds
corresponding to a number of current experiments. Moreover, and maybe
equally important, it provides physical insight into the plasma dynamics
since it is based on a few simple evolution equations for the macroscopic
observables characterizing the state of the plasma. As we have investigated
in detail in \cite{PPR04}, there is generally surprisingly good agreement
between the hydrodynamical simulation and the more sophisticated HMD
calculation as long as macroscopic, i.e.\ spatially averaged, quantities such
as electronic temperature, expansion velocity, ionic correlation energy etc.\
are considered. 
\begin{figure}[tb]
\centerline{\psfig{figure=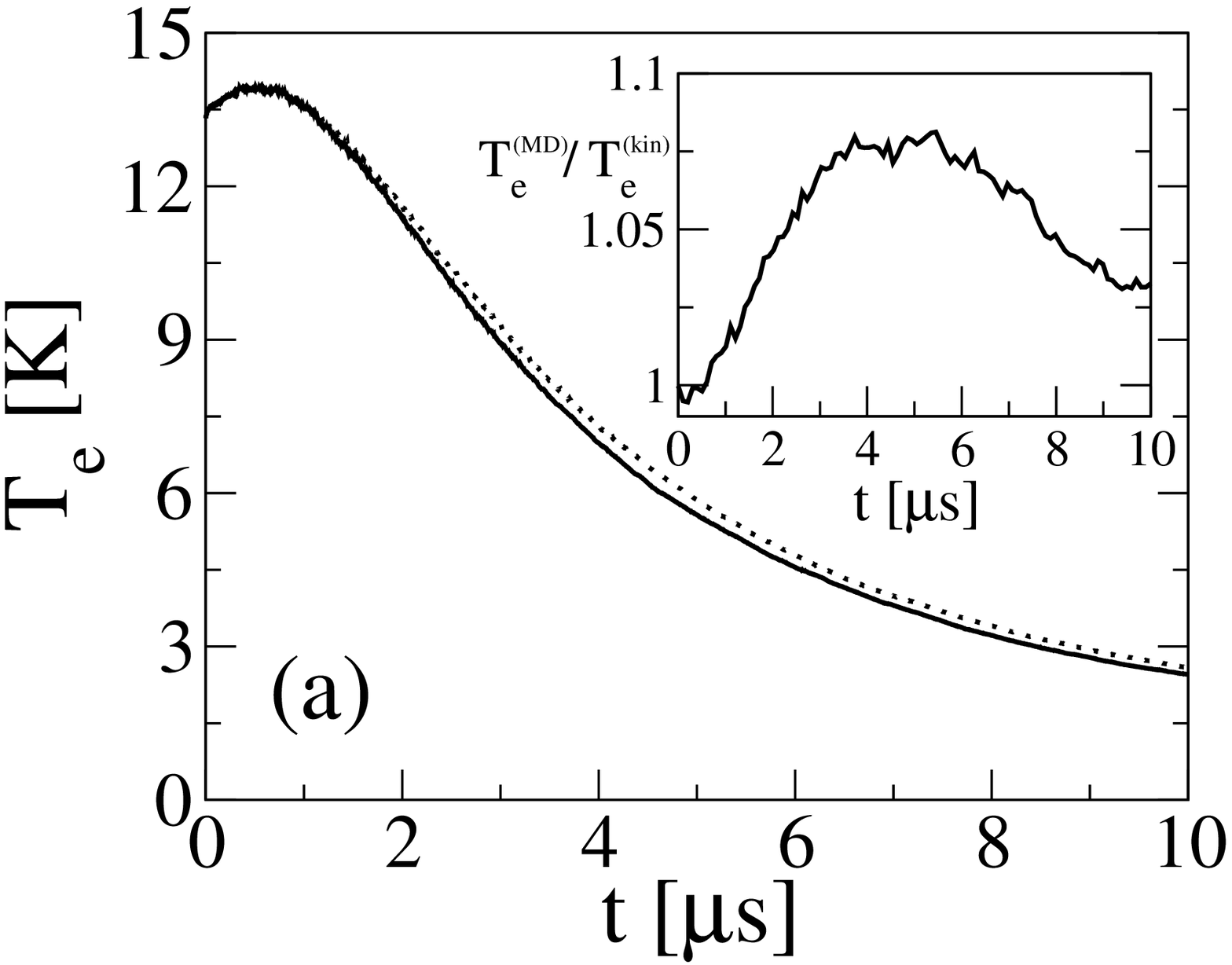,width=6.3cm} \hfill
\psfig{figure=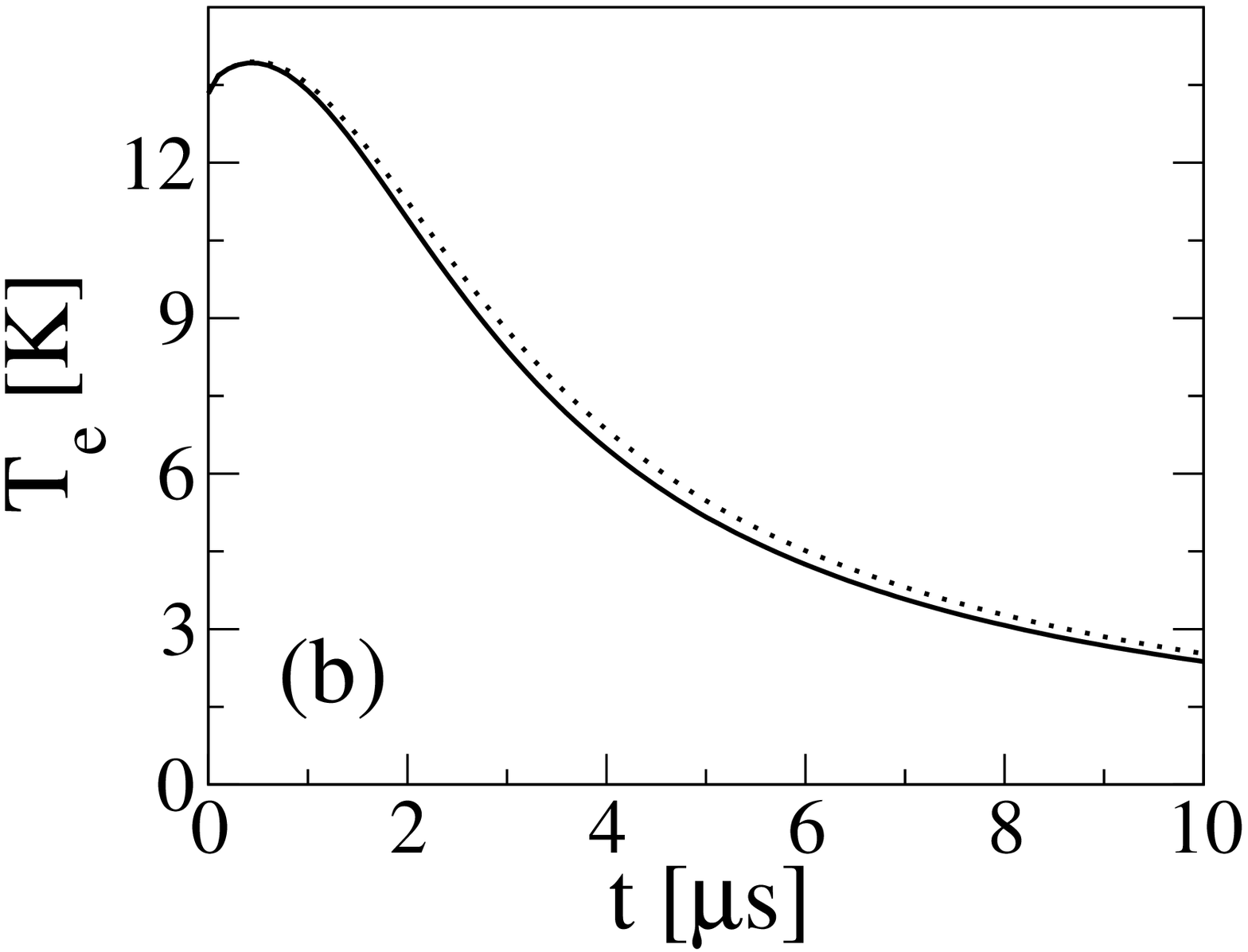,width=6.3cm}}
\caption{\label{f1}
Electronic temperature $T_{\rm{e}}(t)$ for an expanding plasma of $40000$ Sr ions
with an initial average density of $\rho_{\rm{i}}=10^9$cm$^{-3}$ and an initial electron
kinetic energy of $20\:$K, obtained from the HMD simulation (a) and from
eqs.\ (\ref{e2}) (b), with (solid) and without (dotted) the inclusion of ionic
correlations. The inset shows the ratio of the electron temperatures obtained
from both methods.}
\end{figure}
As an example, we show in figure \ref{f1} the time evolution of
the electronic temperature for a plasma of 40000 Sr ions
with an initial average density of $10^9$cm$^{-3}$ and an initial electron
kinetic energy of $20\:$K, obtained from the HMD simulation (a) and from
eqs.\ (\ref{e2}) (b). During the whole system evolution, the agreement between
the two simulation methods is better than about 8\%, and it becomes even
better at later times. Thus, we conclude that, for the present type of
experimental setups, the hydrodynamical method outlined above, and in particular
the approximate treatment of ionic correlations, is well suited for the
description of the behavior of UNPs.

\section{Results and discussion}
\subsection{Comparison with experiments}
\begin{figure}[tb]
\centerline{\psfig{figure=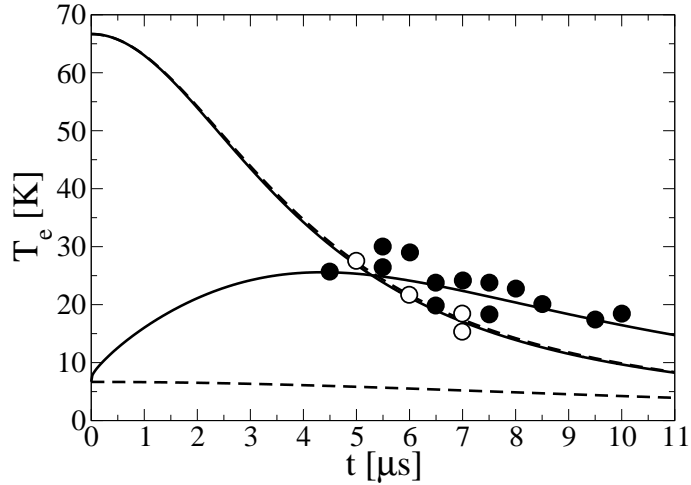,width=9cm}}
\caption{\label{f2}
Electronic temperature $T_{\rm{e}}(t)$ for a plasma of $1.2\cdot10^6$ Xenon ions with
an initial average density of $1.35\cdot10^9$cm$^{-3}$ for two different
initial electron temperatures, $T_{\rm{e}} =6.67$K (filled dots) and $T_{\rm{e}} = 66.67$K
(open dots). The lines show the hydrodynamical simulation (solid lines:
including inelastic collisions, dashed lines: without inelastic collisions),
the dots the experiment \cite{Rob04}, scaled down by 26\% (see text).}
\end{figure}

In fact, fig.\ \ref{f1} only shows good agreement between the two theoretical
simulation methods, without comparison with experiment. Such a comparison is
now also possible, since measurements of the electron temperature dynamics have
recently been reported in \cite{Rob04}. Fig.\ \ref{f2} shows the time evolution
of the electronic temperature for a Xenon plasma with $N_{\rm{i}}(0)=1.2\cdot10^6$,
$\rho_{\rm{i}}(0)=1.35\cdot10^9$cm$^{-3}$ and two different initial temperatures
of $T_{\rm{e}}(0)=66.67$K and $T_{\rm{e}}(0)=6.67$K.
In addition to the full hydrodynamical simulation according to equations
(\ref{e2}), fig.\ \ref{f2} also shows corresponding calculations where the
effect of inelastic electron-ion collisions, eq.\ (\ref{e2e}), is neglected
(dashed lines). (The plasmas in these experiments are too large to be simulated
using the HMD approach.)
\begin{figure}[tb]
\centerline{\psfig{figure=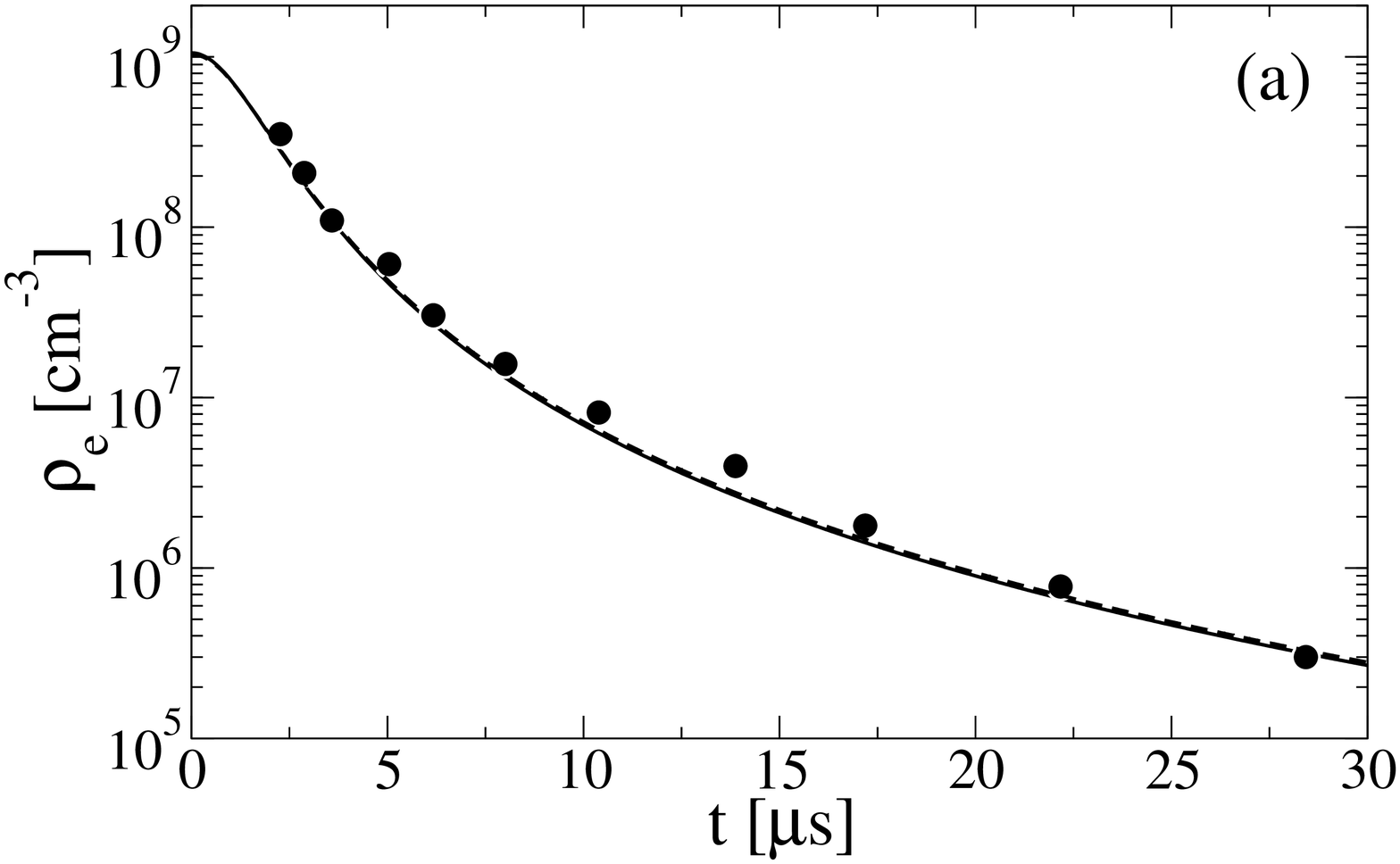,height=3.8cm} \hfill
\psfig{figure=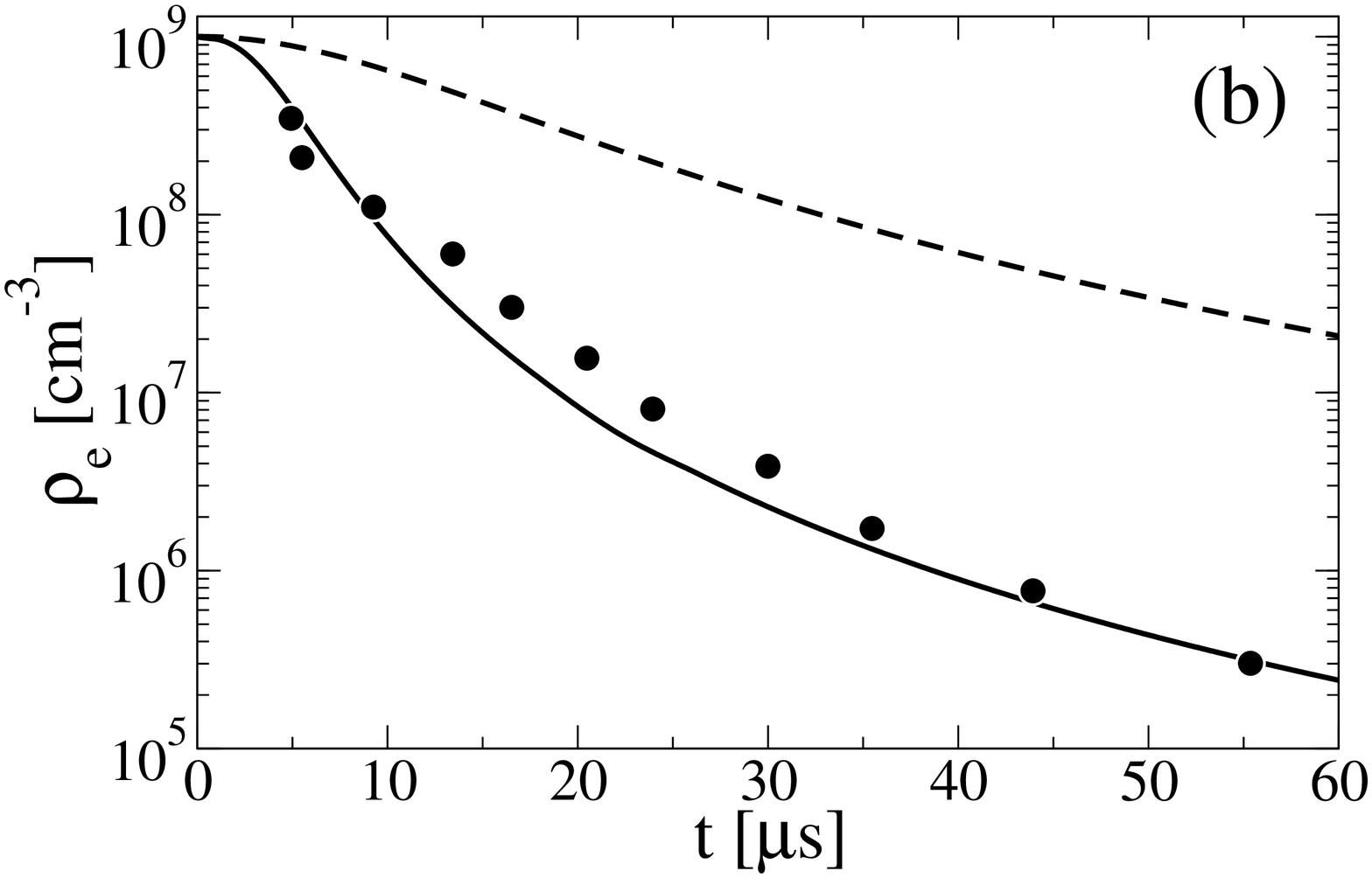,height=3.8cm}}
\caption{\label{f3}
Time evolution of the average electron density of a Xenon plasma of 500000
ions with an initial average density of $10^9$cm$^{-3}$ and an initial electron
temperature of $T_{\rm{e}}=210$K (a) and $T_{\rm{e}}=2.6$K (b). The lines show the results of
the model equations (\ref{e2}) (solid lines: including inelastic collisions,
dashed lines: without inelastic collisions) and the dots the experimental data
from \cite{Kul00}.}
\end{figure}
For the high initial temperature, there is close agreement between
the two corresponding simulations, showing that inelastic processes are almost
negligible in this case. Indeed, it is known that the high-temperature
plasma expansion is well described by the collisionless plasma dynamics, and
the hydrodynamical model is expected to accurately reproduce the plasma
dynamics in this regime. Since an overall systematic error of about $70\%$ for
the temperature measurement has been reported in \cite{Rob04}, we have
exploited this fact to calibrate the measured temperatures by scaling down
both experimental data sets by $26\%$ in order to match the high-temperature
results to our calculations. As can be seen in the figure, there is excellent
agreement between simulation and experiment also for the lower temperature.
(We stress that there is no further scaling of the low-temperature data in order
to achieve quantitative agreement, the same calibration factor as in the
high-temperature case is used.) In this case, inelastic collisions play a
decisive role for the evolution of the system. More specifically, as has been
found already in \cite{Rob02}, three-body recombination heats the plasma and
significantly changes its behavior, leading to a weakly coupled electron gas,
as discussed above in connection with the omission of
electronic correlation effects in the numerical treatment. Moreover, there
has been some discussion in the literature whether the collision rates of
\cite{Man69} would still be applicable at these ultralow temperatures, or
whether three-body recombination would be significantly altered. The close
agreement between the present simulation and the experimental data in fig.\
\ref{f2} suggests that the rates of \cite{Man69}, while ultimately diverging
$\propto T_{\rm{e}}^{-9/2}$ for $T_{\rm{e}} \to 0$, still adequately describe three-body
recombination processes in the temperature range under consideration.

As a second example, figure \ref{f3} shows the time evolution of the electronic
density for a Xenon plasma of $500000$ ions with an initial average density of
$10^9$cm$^{-3}$ and two different initial electron temperatures of
$T_{\rm{e}}(0)=210$K and $T_{\rm{e}}(0)=2.6$K \cite{Kul00}. Again, it can be seen that the
model
equations nicely reproduce the density evolution in both temperature regimes,
in agreement with \cite{Rob02} where it was shown that the low-temperature
enhancement of the expansion velocity \cite{Kul00} is caused by recombination
heating and is not due to strong-coupling effects of the electrons.

\subsection{Role of ionic correlations}
Having thus established the validity of our numerical methods for the
description of UNPs, we can now turn to a more detailed investigation of the
role of ionic correlations in these systems. It is found that, for situations
corresponding to the type of experiments \cite{Kil99,Rob04,Van04}, they
hardly influence the macroscopic expansion behavior of the plasma. This becomes
evident, e.g., in fig.\ \ref{f1}, where the ``full'' simulations as
described above (solid lines) are compared to a mean-field treatment of the
system completely neglecting correlation effects (dotted lines). The
correlation-induced heating of the ions \cite{Mur01,Ger03a,Ger03b} leads
to a slightly faster expansion of the plasma, which in turn results in a
slightly faster adiabatic cooling of the electrons \cite{PPR04}. However, the
overall effect is almost negligible.

\begin{figure}[bt]
\centerline{\psfig{figure=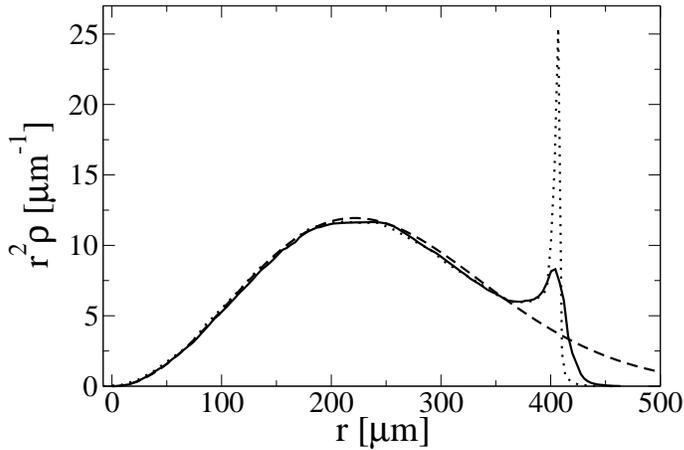,width=9cm}}
\caption{\label{f4}
Spatial density $\rho_{\rm{i}}$ (solid) of the ions, at $t=3\:\mu$s, compared to the
Gaussian profile assumed for the kinetic model (dashed). Additionally, $\rho_{\rm{i}}$
obtained from the particle simulation using the mean-field interaction only is
shown as the dotted line. Initial-state parameters are the same as in fig.\
\ref{f1}.}
\end{figure}
A closer look, on the other hand, reveals that certain aspects of the
expansion dynamics are indeed significantly affected by the strong ion-ion
interaction, as can be seen in figure \ref{f4}. There, the spatial density
of the ions is shown after $t=3\:\mu$s for the same plasma as in fig.\
\ref{f1}. A mean-field treatment of the particle interactions \cite{Rob03}
predicts that a shock front should form at the plasma edge, seen as the
sharp spike in fig.\ \ref{f4} (dotted line). Apparently, with ionic
correlations included (solid line) the peak structure is much
less pronounced than in mean-field approximation. This is due
to dissipation caused by ion-ion collisions which are fully taken into account
in the HMD simulation. As shown in \cite{Sac85}, by adding an ion viscosity
term to the hydrodynamic equations of motion, dissipation tends to
stabilize the ion density and prevents the occurrence of wavebreaking which was
found to be responsible for the diverging ion
density at the plasma edge in the case of a dissipationless plasma
expansion. Furthermore, the initial correlation heating of the ions largely
increases the thermal ion velocities, leading to a broadening of the peak
structure compared to the zero-temperature case.

\begin{figure}[bt]
\centerline{\psfig{figure=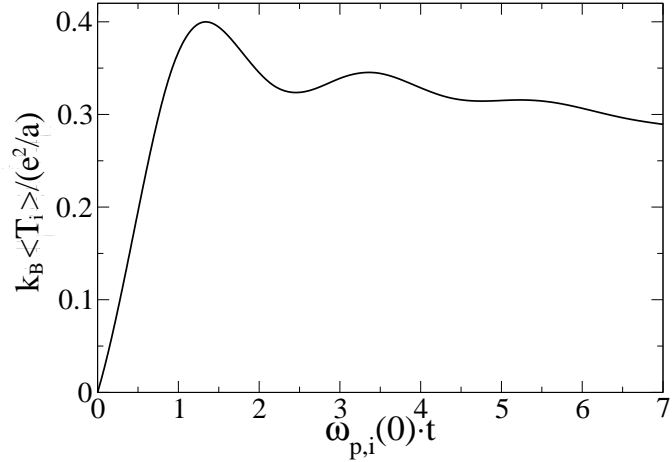,width=9cm}}
\caption{\label{f5}
Time evolution of the density-scaled average ionic temperature for a plasma
consisting of $400000$ ions with an initial electronic Coulomb coupling
parameter of $\Gamma_{\rm{e}}(0)=0.07$.}
\end{figure}

Another obvious aspect where ionic correlations play a dominant role is the
behavior of the ionic temperature. 
Considering the huge ionic coupling
constants suggested by the low initial ion temperatures, this temperature
turns out to be an important quantity since it directly determines the value
of $\Gamma_{\rm{i}}$. According to a mean-field treatment, the ions would
remain the (near) zero temperature fluid they are initially.
However, as has been pointed
out before, the ions are created in a completely uncorrelated non-equilibrium
state, and they quickly heat up through the build-up of correlations as the
system relaxes toward thermodynamical equilibrium. As shown in
\cite{PPR04}, even at early times the ionic velocity distribution is locally
well described by a Maxwell distribution corresponding to a (spatially) local
temperature, justifying the definition of a --- due to the spherical symmetry
of the plasma --- radius-dependent ion temperature
$T_i(r,t)$. Moreover, if the spatially averaged temperature is
identified with the ion temperature determined by the model equations
(\ref{e2}) one can find again good agreement between both approaches concerning
the timescale of the initial heating as well as the magnitude of the ion
temperature, even at later times \cite{PPR04}. However, as becomes apparent
from fig.\ \ref{f5}, the HMD simulations show temporal oscillations of the
ionic temperature, which can, of course, not be described by the linear ansatz
of the correlation-time approximation. 
Such temporal oscillations of the temperature during the initial relaxation
stage are known from molecular dynamics simulations of homogeneous
one-component \cite{Zwi99} and two-component \cite{Mor03} plasmas, which are
clearly caused by the strongly coupled collective ion dynamics, since they
increase in strength with increasing $\Gamma_{\rm{i}}$ and
disappear for $\Gamma_{\rm{i}}(0)<0.5$ \cite{Zwi99}. 
\begin{figure}[tb]
\centerline{\psfig{figure=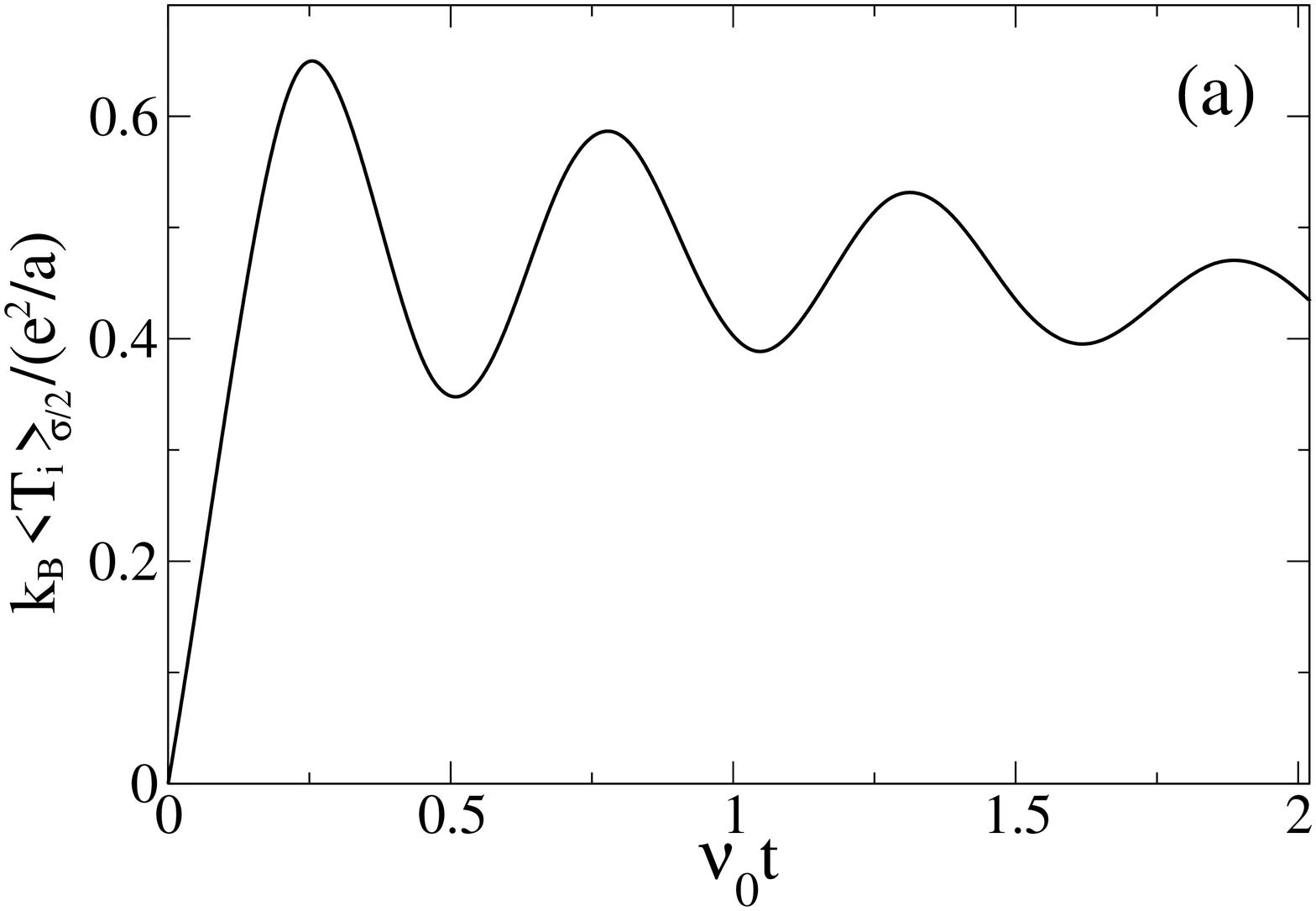,width=6.1cm} \hfill
\psfig{figure=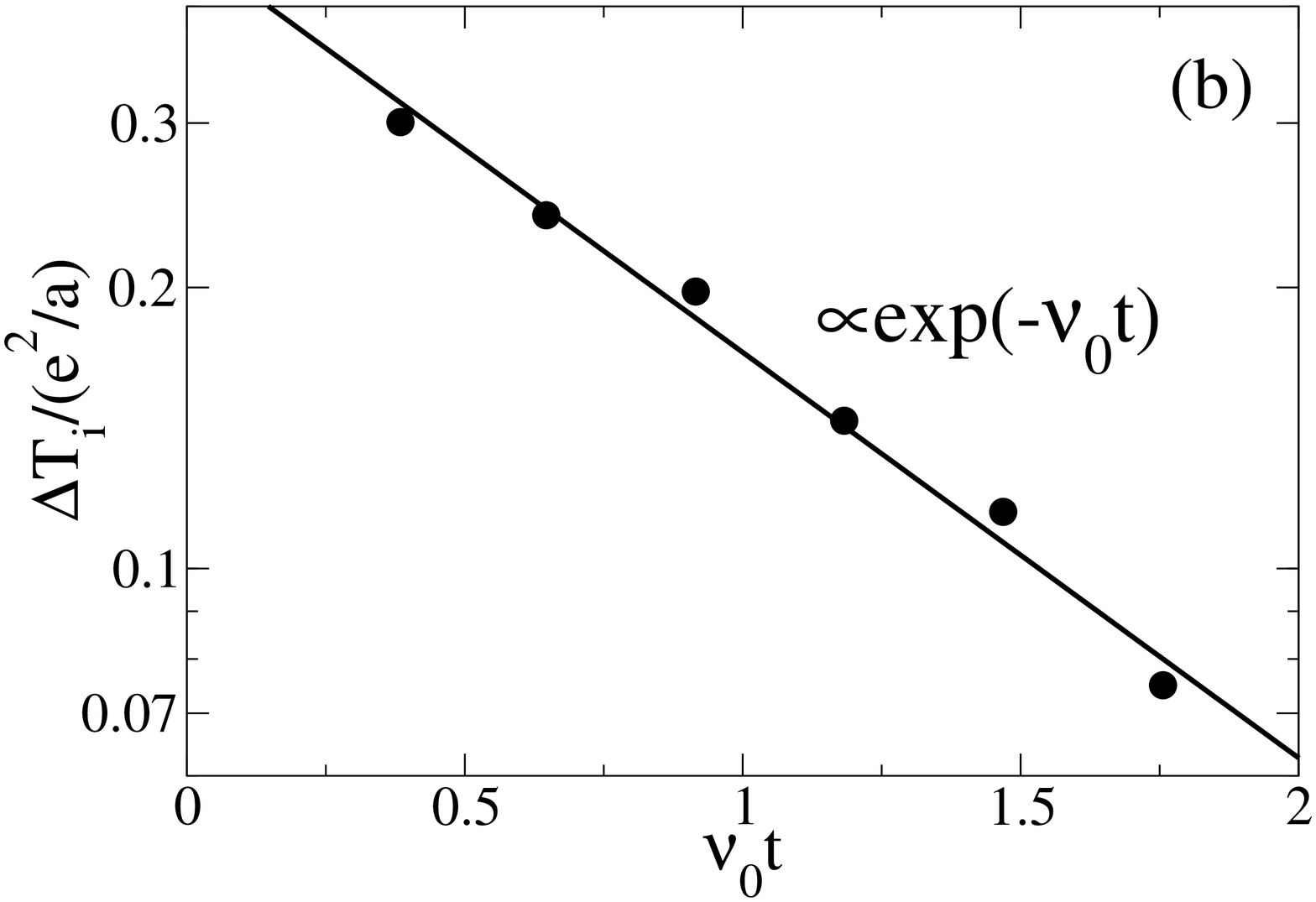,width=6.1cm}}
\caption{\label{f6}
Time evolution of the ion temperature determined from a central sphere
with a radius of half of the plasma width $\sigma$ (a) and time dependence of
the amplitude of the corresponding oscillations (b). The initial-state
parameters are the same as in fig.\ \ref{f5}.}
\end{figure}
\begin{figure}[t]
\centerline{\psfig{figure=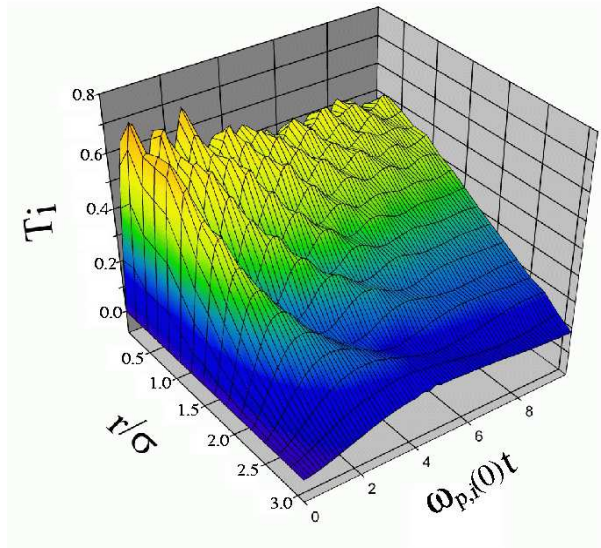,width=8cm}}
\caption{\label{f7}
Temporally and spatially resolved time evolution of the ion temperature. The
initial-state parameters are the same as in fig.\ \ref{f5}.}
\end{figure}

Despite the fact that the maximum initial coupling constant used in
\cite{Zwi99} is $\Gamma_{\rm{i}}=5$, while a value of $\Gamma_{\rm{i}}(0)\approx40000$ is
considered in the case of fig.\ \ref{f5}, the
oscillations observed in \cite{Zwi99} are much more pronounced and persist much
longer than in the present case. It becomes apparent that the rapid damping of
the temperature oscillations can be traced to the inhomogeneity of the Gaussian
density profile by looking at the central part of the plasma only, where the
ionic density is approximately constant (figure \ref{f6}).
The temperature oscillations with an oscillation period
of half of the inverse plasma frequency $\nu_0=\omega_{{\rm{p,i}}}(0)/2\pi$ defined
in the central plasma region are much more pronounced in this case, showing
an exponential decay with a characteristic damping rate of $\nu_0$ (fig.\
\ref{f6}(b)). The temporally and spatially resolved temperature evolution shown
in fig.\ \ref{f7} shows that the radially decreasing ion density leads to
local temperature oscillations with radially increasing frequencies, thereby
causing also spatial oscillations of the local ion temperature.  Therefore,
the seemingly enhanced damping rate, which has also been observed in recent
experiments, is purely an effect of the averaging of these local oscillations
over the total plasma volume.

\subsection{Coulomb crystallization through laser cooling}
The above considerations show that, while not dramatically affecting the
overall expansion behavior of the plasma cloud, strong-coupling effects
play an important role in different aspects of the evolution of UNPs. Thus,
UNPs provide a prime example of laboratory realizations of strongly nonideal
plasmas. Moreover, the HMD approach developed in \cite{PPR04} is well suited
for an accurate description of these systems over experimentally relevant
timescales, allowing for direct comparison between experiment and theory.
Many interesting aspects of the relaxation behavior of these non-equilibrium
plasmas may thus be studied in great detail. However, while effects of strong
ionic coupling become apparent in UNPs, 
the naively expected
regime with $\Gamma > 100$ can not be reached with
the current experimental setups. For the electrons, it is predominantly
three-body recombination which heats them by several orders of magnitude, so
that $\Gamma_{\rm{e}} < 0.2$ during the whole system evolution. The ionic component,
on the other hand, is heated by the correlation-induced heating until
$\Gamma_{\rm{i}} \approx 1$, i.e.\ just at the border of the strongly coupled regime
\cite{Mur01,Sim04}. Thus, it is the very build-up of ionic correlations one
wishes to study that eventually shuts off the process and limits the amount
of coupling achievable in these systems.
\begin{figure}[bt]
\centerline{\psfig{figure=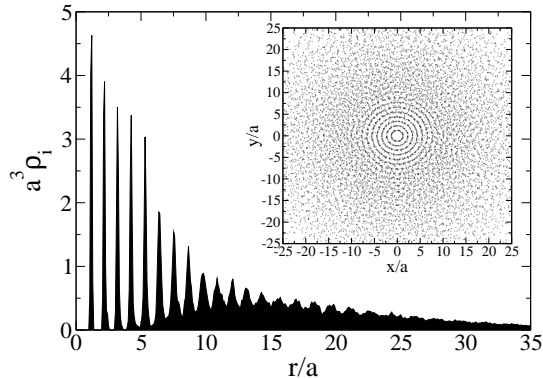,height=5cm}}
\caption{\label{f8}
Radial density and a central slice of a plasma with $N_{\rm{i}}(0)=80000$,
$\Gamma_{\rm{e}}(0)=0.08$, cooled with a damping rate of
$\beta=0.2\omega_{\rm{p,i}}(0)$ at a time of $\omega_{\rm{p,i}}(0)t=216$. (For
better contrast, different cuts have been overlayed.)}
\end{figure}

As soon as the reason for this ionic heating became clear, several proposals
have been made in order to avoid or at least reduce the effect, among them
({\em i}) using fermionic atoms cooled below the Fermi temperature in the
initial state, so that the Fermi hole around each atom prevents the occurrence
of small interatomic distances \cite{Mur01}; ({\em ii}) an intermediate step
of exciting atoms into high Rydberg states, so that the interatomic spacing is
at least twice the radius of the corresponding Rydberg state \cite{Ger03a};
and ({\em iii}) the continuous laser-cooling of the plasma ions after their
initial creation, so that the correlation heating is counterbalanced by the
external cooling \cite{Kil03,PPR04a}. We have simulated the latter scenario
using the HMD method, extended to allow for the description of laser cooling,
as well as elastic electron-ion collisions which are negligible for the free
plasma expansion but not necessarily in the laser-cooled case
\cite{PPR04a,PPR04c}. Laser cooling is modeled by adding a Langevin force
\begin{equation}
{\bf{F}}_{\rm cool} =-m_i\beta{\bf{v}}+\sqrt{2\beta k_{\rm{B}}T_c m_i}
{\bm{\xi}}
\end{equation}
to the ion equation of motion, where ${\bf{v}}$ is the ion
velocity, ${\bm{\xi}}$ is a stochastic variable with $\left<{\bm{\xi}}\right>
={\bf{0}}$, $\left<{\bm{\xi}}(t){\bm{\xi}}(t+\tau)\right>=3\delta(\tau)$,
and the cooling rate $\beta$ and the
corresponding Doppler temperature $T_c$ are determined by the
properties of the cooling laser \cite{Met99}. Elastic electron-ion collisions
are taken into account on the basis of the corresponding Boltzmann collision
integral, which is again evaluated by a Monte-Carlo procedure \cite{PPR04c}. 
\begin{figure}[tb]
\centerline{\psfig{figure=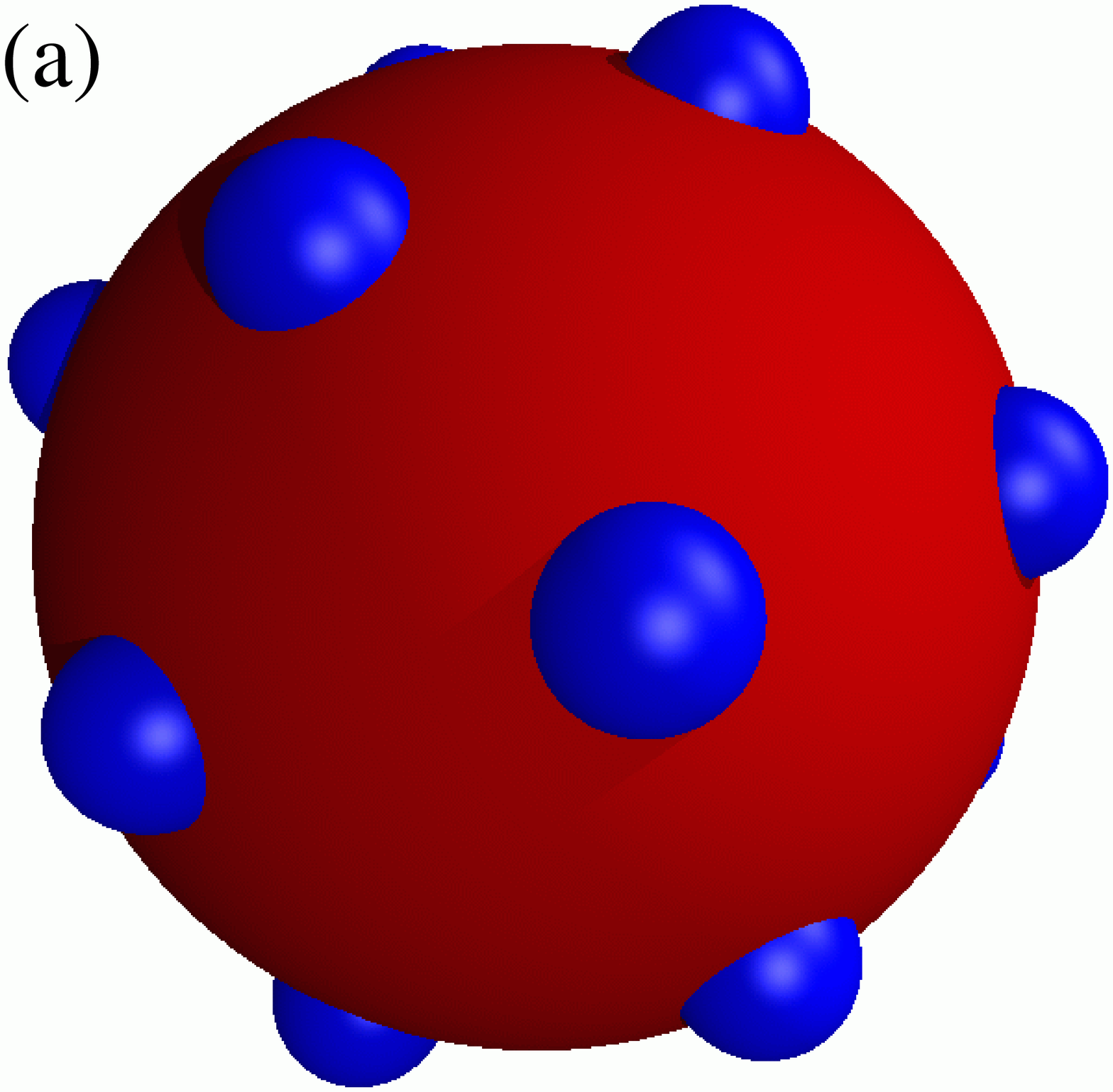,width=3.8cm} \hfill
\psfig{figure=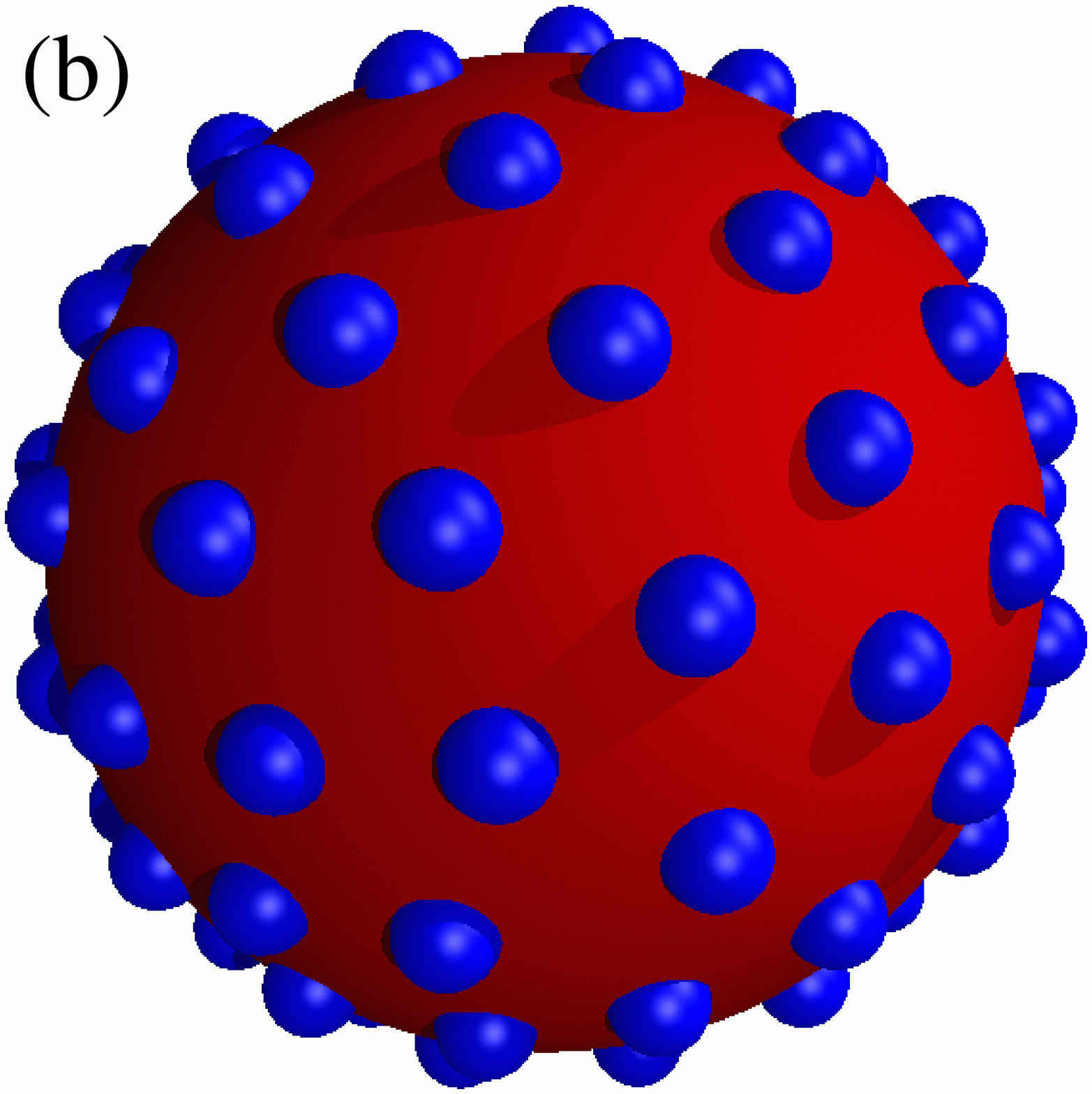,width=3.7cm} \hfill \psfig{figure=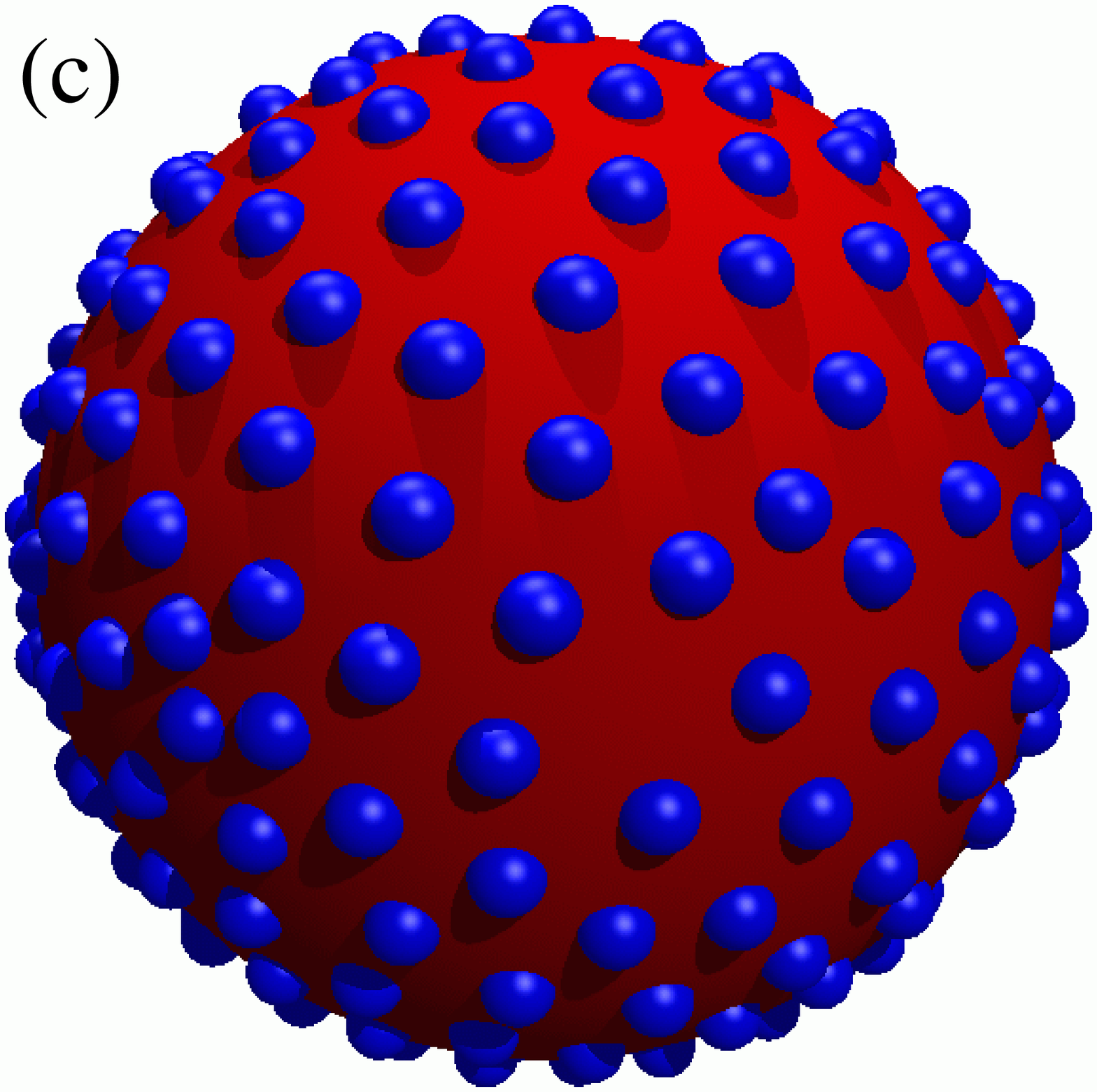,width=3.7cm}}
\caption{\label{f9}
Arrangement of the ions on the first (a), third (b) and fifth (c)
shell of the plasma of fig.\ \ref{f8}.}
\end{figure}

It is found that laser cooling leads to qualitative changes of the plasma
dynamics. In particular, it significantly decelerates the expansion of the
plasma, whose width is found to increase only as $\sigma\propto t^{1/4}$, in
contrast to freely expanding plasmas which behave as $\sigma\propto t$. It is
this drastic slow-down of the expansion which favors the development of strong
ion correlations, compared to a free plasma where the expansion
considerably disturbs the relaxation of the system. The simulations show
further that
strongly coupled expanding plasmas can indeed be created under realistic
conditions, with ionic coupling constants far above the crystallization limit
for homogeneous plasmas of $\Gamma_{\rm{i}}\approx174$ \cite{Dub99}. Here we
find, depending on the initial conditions, i.e.\ ion number
and initial electronic Coulomb coupling parameter, strong
liquid-like short-range correlations or even the onset of a radial
crystallization of the ions. This is demonstrated in fig.\ \ref{f8}, showing the
radial density and a central slice of a plasma with $N_{\rm{i}}(0)=80000$,
$\Gamma_{\rm{e}}(0)=0.08$, cooled with a damping rate of
$\beta=0.2\omega_{\rm{p,i}}(0)$, at a scaled time of
$\omega_{\rm{p,i}}(0)t=216$. The formation of concentric shells in the center of the cloud is clearly visible. As
illustrated in fig.\ \ref{f9}, beside
the radial ordering there is also significant intra-shell ordering, namely a formation of hexagonal structures on the
shells, which are, however, considerably disturbed by the curvature of the
shells.

\section{Conclusions}
In summary, we have used an HMD approach to study the behavior
of ultracold neutral plasmas on long time scales. We have shown that effects of
strong interionic
coupling are indeed visible in such systems, e.g.\ most prominently in the
relaxation behavior of the ion temperature, which is connected
with transient temporal as well as spatial oscillations.
Nevertheless, the strongly coupled regime of $\Gamma > 100$ is not
reached with the current experimental setups. We have demonstrated, however,
that additional continuous laser cooling of the ions during the plasma
evolution qualitatively changes the expansion behavior of the system and
should allow for the Coulomb crystallization of the plasma
\cite{PPR04a,PPR04c}. It will be an interesting subject for further
investigation to study in detail the dynamics of this
crystallization process, which differs from the shell structure formation
observed in trapped nonneutral plasmas \cite{Dub99} as explained in
\cite{PPR04a}. In particular, the
influence of the plasma expansion, which presumably causes the transition from
liquid-like short-range correlation to the radial ordering, deserves more
detailed studies.
Other future directions include the study of effects
induced by additional magnetic fields, or of ways to confine the plasma in
a trap.

We gratefully acknowledge many helpful discussions with J.M.\ Rost, as well as
conversations with T.C.\ Killian and F.\ Robicheaux.


\begin{thebibliography}{30}
\bibitem{Kil99} T.C.\ Killian, S.\ Kulin, S.D.\ Bergeson, L.A.\ Orozco,
C.\ Orzel and S.L.\ Rolston, Phys.\ Rev.\ Lett.\ {\bf 83}, 4776 (1999).

\bibitem{Kul00} S.\ Kulin, T.C.\ Killian, S.D.\ Bergeson and S.L.\ Rolston,
Phys.\ Rev.\ Lett.\ {\bf 85}, 318 (2000).

\bibitem{Sim04} C.E.\ Simien, Y.C.\ Chen, P.\ Gupta, S.\ Laha, Y.N.\ Martinez,
P.G.\ Mickelson, S.B.\ Nagel and T.C.\ Killian, Phys.\ Rev.\ Lett.\ {\bf 92},
143001 (2004).

\bibitem{Kil01} T.C.\ Killian, M.J.\ Lim, S.\ Kulin, R.\ Dumke, S.D.\ Bergeson
and S.L.\ Rolston, Phys.\ Rev.\ Lett.\ {\bf 86}, 3759 (2001).

\bibitem{Rob04} J.L.\ Roberts, C.D.\ Fertig, M.J.\ Lim and S.L.\ Rolston,
Phys.\ Rev.\ Lett.\ {\bf 92}, 253003 (2004).

\bibitem{Van04} N.\ Vanhaecke, D.\ Comparat, D.A.\ Tate and P.\ Pillet, arXiv:physics/0401045

\bibitem{Kli72} Yu.L. Klimontovich, Sov. Phys. JETP {\bf 35}, 920 (1972).

\bibitem{Kli73} Yu.L. Klimontovich and W. Ebeling, Sov. Phys. JETP {\bf 36}, 476 (1973).

\bibitem{Kli82} Yu.L. Klimontovich, {\it{Kinetic theory of nonideal gases and
nonideal plasmas}} (Pergamon Press, 1982).

\bibitem{Wal78} J. Wallenborn and M. Baus, Phys. Rev. A {\bf 18}, 1737 (1978).

\bibitem{Bel96} V.V. Belyi, Yu.A. Kukharenko and J. Wallenborn,
Phys. Rev. Lett. {\bf 76}, 3554 (1996).

\bibitem{Bon98} M. Bonitz {\it{Quantum Kinetic Theory}} (Teubner, 1998).

\bibitem{Zwi99} G.\ Zwicknagel, Contrib.\ Plasma Phys.\ {\bf 39}, 155 (1999).

\bibitem{PPR04} T.\ Pohl, T.\ Pattard and J.M.\ Rost, arXiv:physics/0405125
(2004).

\bibitem{Hah02} Y.\ Hahn, Phys.\ Lett.\ A {\bf 293}, 266 (2002).

\bibitem{Kuz02} S.G.\ Kuzmin and T.M.\ O'Neil, Phys.\ Rev.\ Lett.\ {\bf 88},
065003 (2002).

\bibitem{Rob02} F.\ Robicheaux and J.D.\ Hanson, Phys.\ Rev.\ Lett.\ {\bf 88},
055002 (2002).

\bibitem{Kin66} I.R.\ King, Astron.\ J.\ {\bf 71}, 64 (1966).

\bibitem{Kli81} Yu.L. Klimontovich and D. Kremp, Physica A {\bf 109}, 517 (1981)
 
\bibitem{Man69} P.\ Mansbach and J.\ Keck, Phys.\ Rev.\ {\bf 181}, 275 (1969).

\bibitem{Rob03} F.\ Robicheaux and J.D.\ Hanson, Phys.\ Plasmas {\bf 10}, 2217
(2003).

\bibitem{PPR04c} T.\ Pohl, T.\ Pattard and J.M.\ Rost, J.\ Phys.\ B accepted
(2004).

\bibitem{Bar90} J.E.\ Barnes, J.\ Comp.\ Phys.\ {\bf 87}, 161 (1990).

\bibitem{PPR04a} T.\ Pohl, T.\ Pattard and J.M.\ Rost, Phys.\ Rev.\ Lett.\ {\bf
92}, 155003 (2004).

\bibitem{PPR04b} T.\ Pohl, T.\ Pattard and J.M.\ Rost, J.\ Phys.\ B {\bf 37},
L183 (2004).

\bibitem{Dor98} D.S.\ Dorozhkina and V.E.\ Semenov, Phys.\ Rev.\ Lett.\ {\bf
81}, 2691 (1998).

\bibitem{Bon96} M.\ Bonitz, Phys.\ Lett.\ A {\bf 221}, 85 (1996).

\bibitem{Cha98} G.\ Chabrier and A.Y.\ Potekhin, Phys.\ Rev.\ E {\bf 58},
4941 (1998).

\bibitem{Mur01} M.S.\ Murillo, Phys.\ Rev.\ Lett.\ {\bf 87}, 115003 (2001).

\bibitem{Ger03a} D.O.\ Gericke and M.S.\ Murillo, Contrib.\ Plasma Phys.\
{\bf 43}, 298 (2003).

\bibitem{Ger03b} D.O. Gericke, M.S. Murillo, D. Semkat, M. Bonitz and
D. Kremp, J. Phys. A {\bf 36} 6087 (2003).

\bibitem{Sac85} C.\ Sack and H.\ Schamel, Plasma\ Phys.\ Contr.\ F. {\bf 27},
717 (1985).

\bibitem{Mor03} I.V. Morozov and G.E. Norman, J. Phys. A {\bf 36}, 6005 (2003).

\bibitem{Kil03} T.C.\ Killian, V.S.\ Ashoka, P.\ Gupta, S.\ Laha, S.B.\ Nagel,
C.E.\ Simien, S.\ Kulin, S.L.\ Rolston and S.D.\ Bergeson, J.\ Phys.\ A
{\bf 36}, 6077 (2003).

\bibitem{Met99} H.J.\ Metcalf and P.\ van der Straten, {\it Laser Cooling and
Trapping} (Springer, 1999).

\bibitem{Dub99} D.H.E. Dubin and T.M. O'Neil, Rev. Mod. Phys. {\bf 71}, 87 (1999)
\end{thebibliography}
\end{document}